\documentclass[12pt,a4paper]{iopart}

\usepackage{graphicx}
\usepackage{dcolumn}
\usepackage{bm}

\begin{document}

\title[Superconductivity in oxygen-added Zr$_{5}$Pt$_{3}$]{Superconductivity in oxygen-added Zr$_{5}$Pt$_{3}$}

\author{Shusuke Hamamoto$^1$ and Jiro Kitagawa$^1$}

\address{$^1$ Department of Electrical Engineering, Faculty of Engineering, Fukuoka Institute of Technology, 3-30-1 Wajiro-higashi, Higashi-ku, Fukuoka 811-0295, Japan}
\ead{j-kitagawa@fit.ac.jp}

\begin{abstract}
Mn$_{5}$Si$_{3}$-type structure has been offering an interstitial chemistry. Recent report of enhancement of superconductivity in a Nb-based Mn$_{5}$Si$_{3}$-type compound by addition of an interstitial atom motivated us to investigate the effect of oxygen-addition in Mn$_{5}$Si$_{3}$-type Zr$_{5}$Pt$_{3}$. The superconducting critical temperature of 6.4 K in the parent Zr$_{5}$Pt$_{3}$ is monotonously reduced to 3.2 K in Zr$_{5}$Pt$_{3}$O$_{x}$ ($x$=0.6) with increasing oxygen-content. As $x$ is further increased from 0.6 to 2.5, exceeding the full occupancy of oxygen site ($x$=1.0), samples become multi-phases composed of Zr$_{5}$Pt$_{3}$O$_{\sim 0.5-0.6}$, ZrPt and ZrO$_{2}$. However, the superconducting critical temperature slightly increases to 4.8 K at $x$=2.5. The metallographic observation has revealed a change of microstructure at $x \geq$1.0. The change of microstructure and/or the composition effect would be responsible for the enhancement of superconductivity.
\end{abstract}

\vspace{2pc}
\noindent{\it Keywords}: Mn$_{5}$Si$_{3}$-type, oxygen addition, superconductivity, microstructure, composition effect

\maketitle

\clearpage

\section{Introduction}
A wide variety of atoms form the hexagonal Mn$_{5}$Si$_{3}$-type structure\cite{Corbett:CM1998}, represented as A$_{5}$B$_{3}$, with the space group P6$_{3}$/mcm (No.193).
There are three crystallographic sites in A$_{5}$B$_{3}$.
The A atoms occupy the 4d (for A1 atom) and 6g (for A2 atom) sites and the B atom occupies another 6g site.
The A atoms mainly consist of early transition metals, rare earth elements and alkaline earth elements.
Metalloid elements and post-transition metals are usually responsible for B atoms.  
The extensive electron-number of A$_{5}$B$_{3}$ allows various interstitial atoms such as oxygen, boron and carbon.
These interstitial atoms, denoted by X, occupy the 2b site in P6$_{3}$/mcm and A$_{5}$B$_{3}$X is called as the Ti$_{5}$Ga$_{4}$ or Hf$_{5}$CuSn$_{3}$-type structure.
Figure 1 shows the crystal structure of A$_{5}$B$_{3}$X compound.
The added X atom is surrounded by A2 atoms in the octahedral site, which forms a face-sharing A2$_{6}$ chain along the $c$-axis.
Another octahedral B atoms enclose the A1 atom which also forms a one-dimensional atomic chain along the $c$-axis.
Therefore an expansion of lattice parameter $a$ would enhance the one-dimensional nature of octahedral A2$_{6}$ and A1 atomic chains.

\begin{figure}
\begin{center}
\includegraphics[width=13cm]{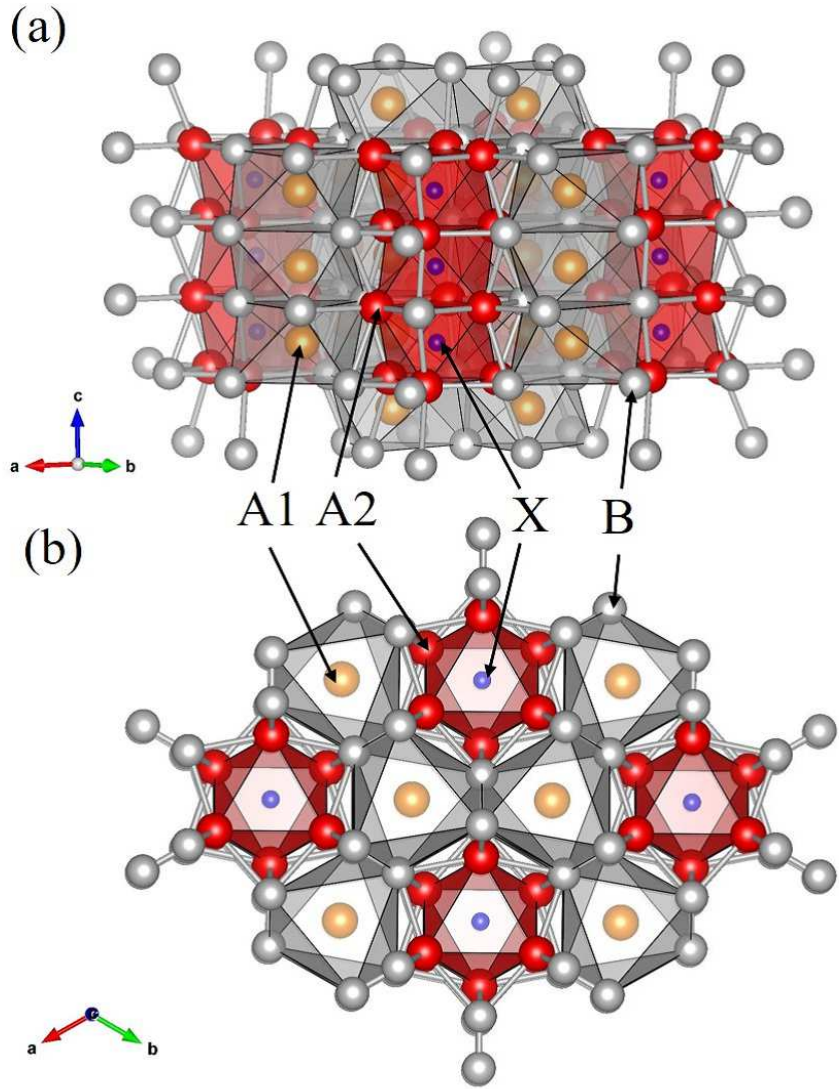}
\end{center}
\caption{Crystal structure of Ti$_{5}$Ga$_{4}$-type A$_{5}$B$_{3}$X compound (a) along the $c$-axis and (b) in the $a$-$b$ plane, respectively.}
\label{f1}
\end{figure}

Although physical properties of numerous Mn$_{5}$Si$_{3}$ or Ti$_{5}$Ga$_{4}$-type compounds are investigated\cite{Zheng:JALCOM2002,Surgers:PRB2003,Mar:CM2006,Goruganti:JAP2009}, the superconductivity is reported only in several compounds.
The Nb-based system Nb$_{5}$Ir$_{3}$O is attractive as a rather high superconducting critical temperature $T_{c}$ compound.
In Nb$_{5}$Ir$_{3}$O, the parent Mn$_{5}$Si$_{3}$-type Nb$_{5}$Ir$_{3}$ exhibits superconductivity at 9.3 K.
As the oxygen atoms are added, $T_{c}$ progressively increases to 10.5 K in Nb$_{5}$Ir$_{3}$O, which possesses two kinds of superconducting gaps revealed by the specific heat measurement\cite{Zhang:npjQM2017}.
  
As for Zr-based Mn$_{5}$Si$_{3}$-type compounds, Lv et al. have reported the superconducting Zr$_{5}$Sb$_{3}$ with $T_{c}$ of 2.3 K\cite{Lv:PRB2013}.
Zr$_{5}$Sb$_{3}$ allows interstitial oxygen atoms fully occupying the 2b site.
Contrary to the enhancement of $T_{c}$ in Nb$_{5}$Ir$_{3}$O, the addition of oxygen atoms reduces $T_{c}$ and Zr$_{5}$Sb$_{3}$O is a normal metal down to 1.8 K\cite{Lv:PRB2013}.
Recently it has been discovered\cite{Li:NJP2018} that a Ru substitution into the Ge site in non-superconducting Zr$_{5}$Ge$_{3}$ induces a superconducting behavior at 5.7 K.
About 30 years ago, Waterstrat et al. reported\cite{Waterstrat:JLCM1990} that Zr$_{5}$Pt$_{3}$ is a superconductor with $T_{c}$ of 7.2 K.
Furthermore, oxygen-added Zr$_{5}$Pt$_{3}$O is reported to crystallize in the Ti$_{5}$Ga$_{4}$-type structure\cite{Gupta:JSSC2009}, however, the physical properties of Zr$_{5}$Pt$_{3}$O have not been investigated.
We have focused on the effect of oxygen-addition in superconducting Zr$_{5}$Pt$_{3}$.
In this paper, we report the synthesis and metallographic characterization of Zr$_{5}$Pt$_{3}$O$_{x}$ and the oxygen-content dependence of $T_{c}$.

\section{Materials and Methods}
Polycrystalline samples were prepared using Zr pieces (or powder) (99\%), Pt wire (99.9\%) and ZrO$_{2}$ powder (98\%).
Zr$_{5}$Pt$_{3}$ was synthesized by arc melting the Zr pieces and Pt wire with the stoichiometric composition.
To synthesize oxygen-added Zr$_{5}$Pt$_{3}$O$_{x}$, Zr$_{5}$O$_{x}$ was initially prepared as follows.
Zr and ZrO$_{2}$ powders were mixed in an agate mortar and pressed into a pellet.
The pelletized sample was arc melted and then Zr$_{5}$O$_{x}$ was remelted with added Pt wire to form the stoichiometric composition.
The samples were remelted several times to ensure the homogeneity of the samples.
The weight loss during the arc melting was negligible.
Each as-cast sample was annealed in an evacuated quartz tube at 800 $^{\circ}$C for 4 days.
The samples were evaluated using a powder X-ray diffractometer (Shimadzu, XRD-7000L) with Cu-K$\alpha$ radiation. 
The metallographic characterization was carried out by observing back-scattered electron images obtained by a field emission scanning electron microscope (FE-SEM; JEOL, JSM-7100F).
The atomic composition of sample was checked by using an energy dispersive X-ray (EDX) spectrometer that was equipped with the FE-SEM. 

The temperature dependence of ac magnetic susceptibility $\chi_{ac}$ (T) in an alternating field of 5 Oe at 800 Hz, between 2.8 K and 300 K, was measured using a closed-cycle He gas cryostat.
The temperature dependence of electrical resistivity $\rho$ (T) between 2.8 K and 300 K was measured by the conventional DC four-probe method using the cryostat.

\section{Results and discussion}
Figure 2(a) shows the X-ray diffraction (XRD) patterns of Zr$_{5}$Pt$_{3}$O$_{x}$ ($x$=0, 0.2, 0.6 and 1.0).
The simulated patterns of Zr$_{5}$Pt$_{3}$ (Mn$_{5}$Si$_{3}$-type) and Zr$_{5}$Pt$_{3}$O (Ti$_{5}$Ga$_{4}$-type) are also presented.
The diffraction peaks of parent compound Zr$_{5}$Pt$_{3}$ and oxygen-added Zr$_{5}$Pt$_{3}$O$_{x}$ with $x$=0.2 and 0.6 samples can be well indexed by the Mn$_{5}$Si$_{3}$-type and the Ti$_{5}$Ga$_{4}$-type structure, respectively.
Although the XRD pattern of Ti$_{5}$Ga$_{4}$-type structure is contained in Zr$_{5}$Pt$_{3}$O, impurity phases of ZrPt (filled triangle) and ZrO$_{2}$ (filled circle) appear.
The ideal uppermost oxygen content is 1.0 by taking into account the full occupancy of 2b site, however, we prepared the samples with $x$ exceeding 1.0.
The peak intensity of impurity phases grows with increasing $x$ from 1.0 (see Fig.\ 2(b)).
In Zr$_{5}$Pt$_{3}$O$_{2.5}$, the maximum peak intensity of ZrPt surpasses that of the Ti$_{5}$Ga$_{4}$-type structure.
The lattice parameters of prepared samples were refined by the least square method using XRD data and listed in Table 1.
The $x$ dependences of $a$ and $c$, and $c/a$-ratio are displayed in Figs.\ 3(a) and 3(b), respectively.
In both figures, the nominal oxygen-content is employed as $x$.
Both $a$ and $c$ systematically decrease with increasing $x$ from 0 to 1.0.
Above $x$=1.0, $c$ steeply increases, while $a$ shows a slight increase.
A monotonous increase of $c/a$-ratio with increasing $x$ is confirmed in Fig.\ 3(b).

\begin{figure}
\begin{center}
\includegraphics[width=12cm]{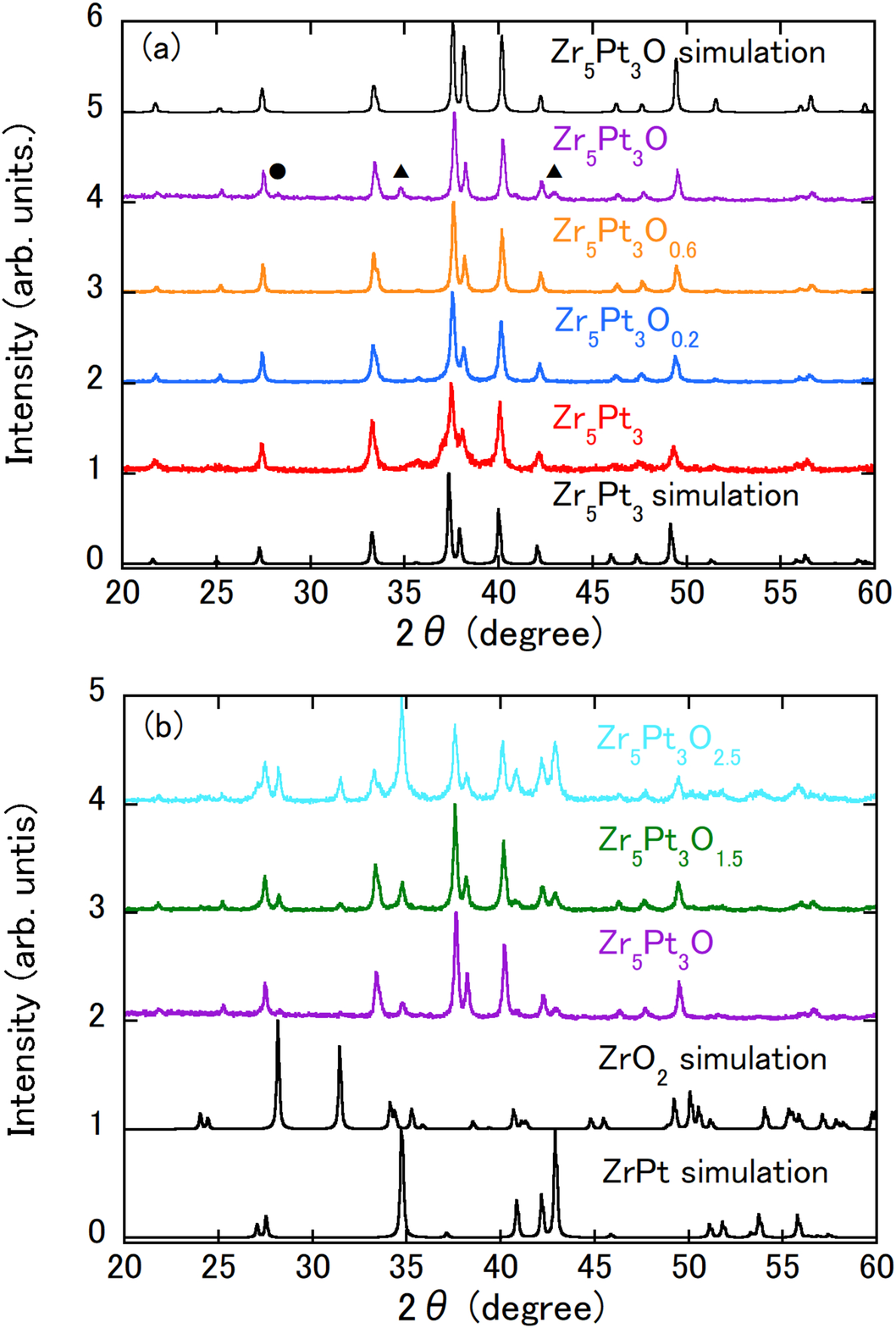}
\end{center}
\caption{(a) XRD patterns of Zr$_{5}$Pt$_{3}$O$_{x}$ with $x$=0, 0.2, 0.6 and 1.0. The simulated patterns of Zr$_{5}$Pt$_{3}$ and Zr$_{5}$Pt$_{3}$O are also shown. The origin of each pattern is shifted by an integer value for clarity. The filled triangle and circle denote impurity phases of ZrPt and ZrO$_{2}$, respectively. (b) XRD patterns of Zr$_{5}$Pt$_{3}$O$_{x}$ with $x$=1.0, 1.5 and 2.5. The simulated patterns of ZrPt and ZrO$_{2}$ are also shown. The origin of each pattern is shifted by an integer value for clarity.}
\label{f2}
\end{figure}

\begin{table}
\caption{Lattice parameters ($a$ and $c$), $c/a$-ratio and $T_{c}$'s determined by $\chi_{ac}$ and $\rho$ measurements for Zr$_{5}$Pt$_{3}$O$_{x}$ and Zr$_{4.9}$Pt$_{3.1}$O$_{0.67}.$}
\label{t1}
\begin{tabular}{cccccc}
\hline
Sample & $a$ (\AA) & $c$ (\AA) & $c/a$ & $T_{c}$ (K) by $\chi_{ac}$ & $T_{c}$ (K) by $\rho$ \\
\hline
Zr$_{5}$Pt$_{3}$ & 8.182(3) & 5.384(2) & 0.6580(5) & 6.4 & 6.4  \\
Zr$_{5}$Pt$_{3}$O$_{0.2}$ & 8.167(2) & 5.375(1) & 0.6581(3) & 4.1 & 4.0  \\
Zr$_{5}$Pt$_{3}$O$_{0.6}$ & 8.156(2) & 5.372(2) & 0.6587(4) & 3.2 & 3.1  \\
Zr$_{5}$Pt$_{3}$O & 8.151(3) & 5.369(2) & 0.6587(5) & 3.5 & 3.2  \\
Zr$_{5}$Pt$_{3}$O$_{1.5}$ & 8.159(3) & 5.376(2) & 0.6590(5) & 4.1 & 3.7  \\
Zr$_{5}$Pt$_{3}$O$_{2.5}$ & 8.157(3) & 5.384(2) & 0.6601(5) & 4.8 & 4.8  \\
Zr$_{4.9}$Pt$_{3.1}$O$_{0.67}$ & 8.159(2) & 5.379(1) & 0.6593(5) & 3.7 & 3.3 \\
\hline
\end{tabular}
\end{table}

\begin{figure}
\begin{center}
\includegraphics[width=12cm]{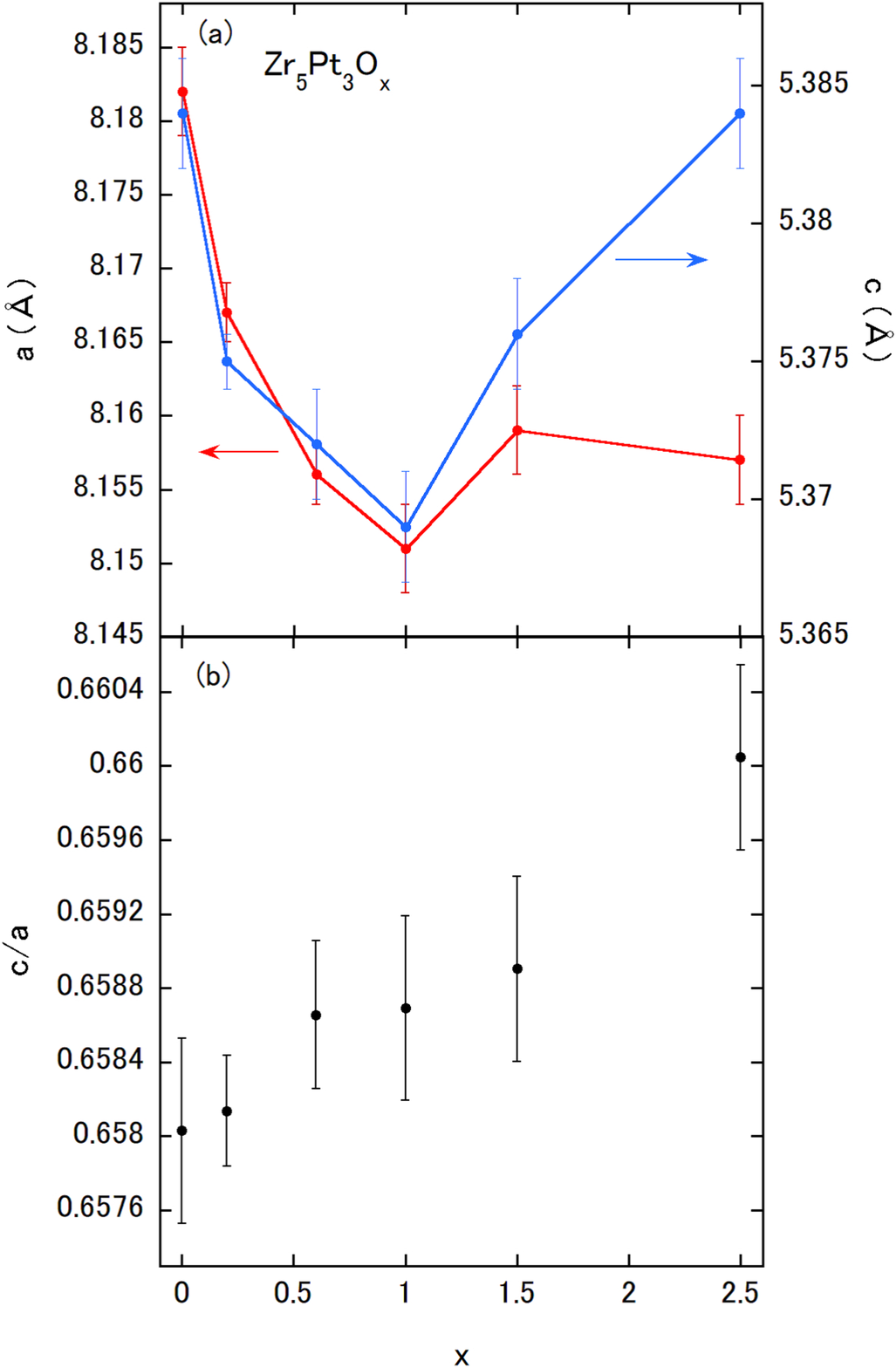}
\end{center}
\caption{(a) $a$ and $c$ versus oxygen-content plots. (b) $c/a$-ratio versus oxygen-content plot. In each figure, the nominal oxygen-content is employed.}
\label{f3}
\end{figure}

Back-scattered electron images obtained by FE-SEM with electron beams of 15 keV are shown in Fig.\ 4.
The atomic composition obtained by EDX measurement of each sample is listed in Table 2.
For each sample with $x\leq$ 0.2, non-contrast image means almost single phase. 
In Zr$_{5}$Pt$_{3}$, Zr$_{5}$Pt$_{3}$O$_{0.2}$ and Zr$_{5}$Pt$_{3}$O$_{0.6}$, the respective sample shows the atomic composition being near to the starting one.
As $x$ is further increased from 1.0, the sample decomposes into three phases Zr$_{5}$Pt$_{3}$O$_{\sim 0.5-0.6}$, ZrPt and ZrO$_{2}$.
Although the composition ratios between Zr and Pt atoms in Zr$_{5}$Pt$_{3}$O$_{x}$ ($x \geq$ 1.0) are close to Zr$_{5}$Pt$_{3}$, the oxygen-contents are far less than the nominal ones, which means a solubility limit of oxygen atoms ($x\sim$ 0.6) in Zr$_{5}$Pt$_{3}$. 
The dark islands observed in Zr$_{5}$Pt$_{3}$O$_{x}$ ($x \geq$1.0) are ZrO$_{2}$ phases, which begin to appear in Zr$_{5}$Pt$_{3}$O$_{0.6}$ as a small amount of point-like structure.
The brighter images in Figs.\ 4(d), 4(e) and 4(f) are ZrPt phases.
It is remarkable that Zr$_{5}$Pt$_{3}$O$_{\sim 0.5-0.6}$ and ZrPt partially forms a eutectic-like structure, for example, denoted by red elliptic closed-curves as in Figs.\ 4(d) and 4(e).
On going from $x$=1.0, 1.5 to 2.5, the area of eutectic-like structure seems to increase.
We note here that it is important to investigate a phase relation, a homogeneity range and so on, for example, using a differential thermal analysis method.
However, even for the binary Zr-Pt system, the phase diagrams in the vicinity of Zr$_{5}$Pt$_{3}$ differ from literature to literature\cite{Biswas:Metallk1967,Massalski:book1990,Alonso:ScrMater2001}, which indicates some difficulties in obtaining a precise phase relation or homogeneity range for Zr$_{5}$Pt$_{3}$O$_{x}$ system.
Therefore, a thermal analysis study would require a long time and careful experiment, and remains an issue.

\begin{figure}
\centering
\includegraphics[width=18cm]{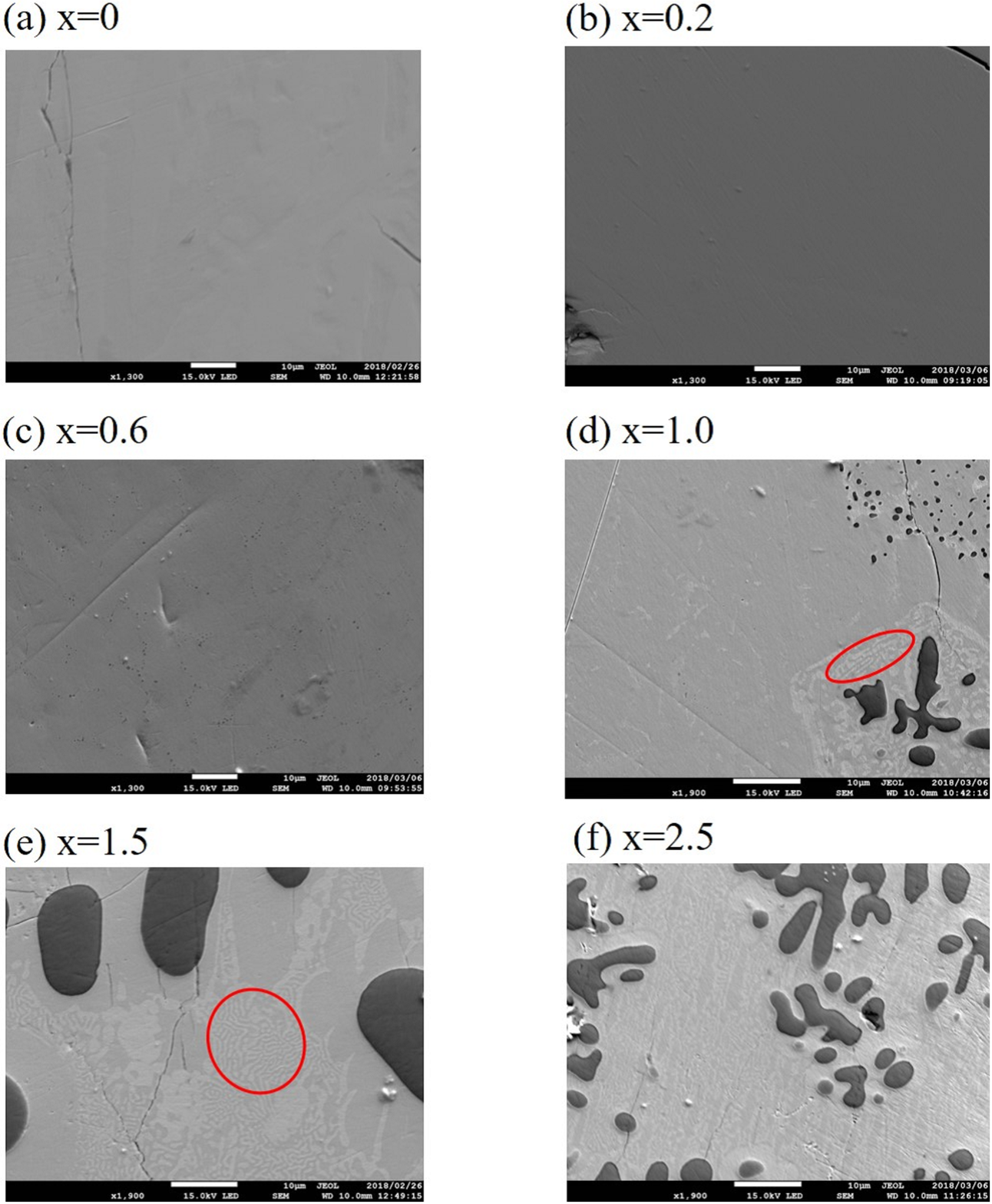}
\caption{Back-scattered electron (15 keV) images of Zr$_{5}$Pt$_{3}$O$_{x}$ for nominal $x$ values of (a) 0 (b) 0.2 (c) 0.6 (d) 1.0 (e) 1.5 and (f) 2.5, respectively.}
\label{f4}
\end{figure}

\begin{table}[t]
\caption{Atomic composition of Zr$_{5}$Pt$_{3}$O$_{x}$ and Zr$_{4.9}$Pt$_{3.1}$O$_{0.67}$ determined by EDX measurement.}
\label{t2}
\begin{tabular}{cc}
\hline
Sample & Atomic composition \\
\hline
Zr$_{5}$Pt$_{3}$ & Zr$_{5.1(2)}$Pt$_{2.9(2)}$  \\
Zr$_{5}$Pt$_{3}$O$_{0.2}$ & Zr$_{4.9(1)}$Pt$_{3.1(1)}$O$_{0.20(9)}$  \\
Zr$_{5}$Pt$_{3}$O$_{0.6}$ & Zr$_{5.1(1)}$Pt$_{2.9(1)}$O$_{0.66(9)}$  \\
Zr$_{5}$Pt$_{3}$O$_{1.0}$ & Zr$_{5.2(2)}$Pt$_{2.8(1)}$O$_{0.45(9)}$, Zr$_{1.0(2)}$Pt$_{1.0(2)}$, Zr$_{1.09(6)}$O$_{1.91(6)}$ \\
Zr$_{5}$Pt$_{3}$O$_{1.5}$ & Zr$_{4.7(2)}$Pt$_{3.2(2)}$O$_{0.55(9)}$, Zr$_{1.0(1)}$Pt$_{1.0(1)}$, Zr$_{1.03(5)}$O$_{1.98(5)}$ \\
Zr$_{5}$Pt$_{3}$O$_{2.5}$ & Zr$_{4.9(1)}$Pt$_{3.1(1)}$O$_{0.67(7)}$, Zr$_{0.99(1)}$Pt$_{1.01(1)}$, Zr$_{0.95(3)}$O$_{2.05(3)}$ \\
Zr$_{4.9}$Pt$_{3.1}$O$_{0.67}$ & Zr$_{4.9(1)}$Pt$_{3.2(2)}$O$_{0.41(7)}$, Zr$_{1.12(6)}$Pt$_{0.88(6)}$ \\
\hline
\end{tabular}
\end{table}

Shown in Figs.\ 5(a) and 5(b) are $\chi_{ac}$ (T) of Zr$_{5}$Pt$_{3}$O$_{x}$ with 0$\leq x \leq$ 0.6 and those with 1.0$\leq x \leq$ 2.5, respectively.
All Zr$_{5}$Pt$_{3}$O$_{x}$ samples exhibit diamagnetic signals.
In each sample, $T_{c}$ was determined as being the intercept of the linearly extrapolated diamagnetic slope with the normal state signal (see the broken lines in the figures), and listed in Table 1.
$T_{c}$ of Zr$_{5}$Pt$_{3}$ is slightly lower than the literature \cite{Waterstrat:JLCM1990} value 7.2 K.
As $x$ is increased from 0 to 0.6, $T_{c}$ systematically decreases to 3.2 K.
However, further increase of $x$ enhances $T_{c}$ up to 4.8 K at $x$=2.5.
As mentioned above, the samples with $x\geq$ 1.0 contain well known insulating ZrO$_{2}$ and ZrPt.
In order to check whether ZrPt is a superconductor or not, $\chi_{ac}$ of ZrPt was measured and no diamagnetic signal down to 2.8 K is observed as shown in Fig.\ 5(b). 
Therefore the observed superconductivities are intrinsic for Zr$_{5}$Pt$_{3}$O$_{\sim 0.5-0.6}$ phases in Zr$_{5}$Pt$_{3}$O$_{x}$ with $x\geq$ 1.0.

\begin{figure}
\begin{center}
\includegraphics[width=12cm]{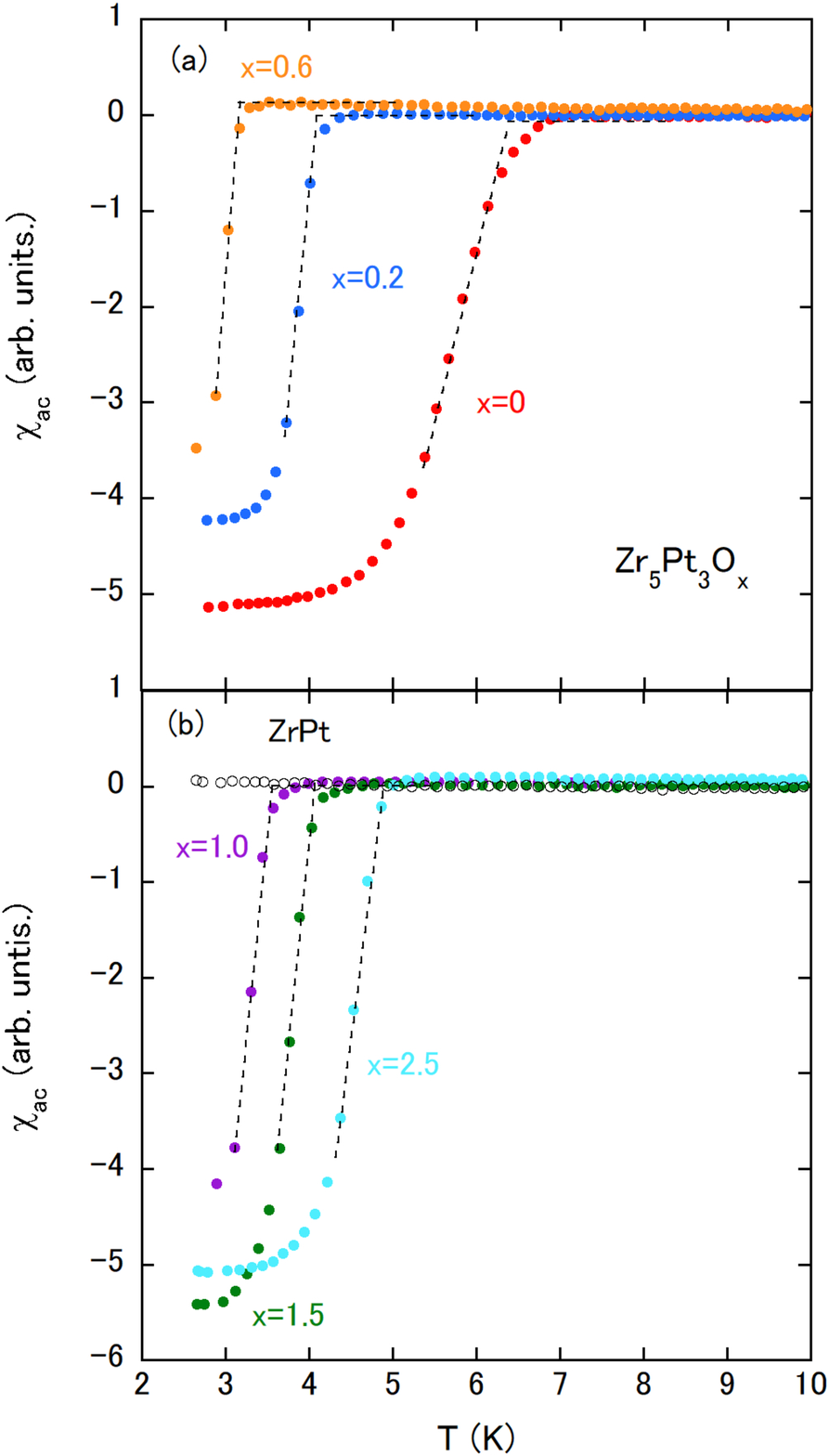}
\end{center}
\caption{(a) Temperature dependence of $\chi_{ac}$ for Zr$_{5}$Pt$_{3}$O$_{x}$ with $x$=0, 0.2 and 0.6. (b) Temperature dependence of $\chi_{ac}$ for Zr$_{5}$Pt$_{3}$O$_{x}$ with $x$=1.0, 1.5 and 2.5. $\chi_{ac}$(T) of ZrPt is also shown.}
\label{f5}
\end{figure}

Figure 6(a) (6(b)) summarizes $\rho$ (T) of Zr$_{5}$Pt$_{3}$O$_{x}$ with 0$\leq x \leq$ 0.6 (1.0$\leq x \leq$ 2.5).
In a few samples with low $T_{c}$, zero resistivity could not be observed at the lowest achievable temperature. 
Each $T_{c}$ determined by following the same manner as in $\chi_{ac}$ (T) roughly corresponds to that obtained by $\chi_{ac}$ (T) measurement (see Table 1).
The sample with $x$=0 (also with 0.2 and 0.6) shows $\rho$ (T) largely deviating from the linearity above $T_{c}$.
The similar deviation is observed in A15 superconductors such as Nb$_{3}$Sn or a pyrochloa superconductor of KOs$_{2}$O$_{6}$, which is ascribed to an existence of additional scattering source\cite{Woodard:PR1964,Hiroi:PRB2007}.
Although the origin of scattering source might be controversial, Woodward and Cody\cite{Woodard:PR1964} have presented a well-known empirical formula as follows:
\begin{equation}
 \rho=\rho_{0}+\rho_{1}T+\rho_{2}exp(-\frac{T_{0}}{T})
\label{equ:WC}
\end{equation}
, where the first term means a residual resistivity, the second one phonon part of $\rho$ and the third one describes anomalous temperature dependence.
In Nb$_{3}$Sn and KOs$_{2}$O$_{6}$, the equation well reproduces respective $\rho$ (T). 
We have also fitted $\rho$ (T) of Zr$_{5}$Pt$_{3}$ using eq. (1).
If the linear term $\rho_{1}T$ is taken into account, the fitting accuracy is not satisfactory.
The well reproducibility as depicted by the solid curve in Fig.\ 6(a) can be obtained by eliminating the linear term, which suggests a dominance of anomalous temperature dependence.
The parameters were determined to be $\rho_{0}=$183 $\mu\Omega$cm, $\rho_{2}=$234 $\mu\Omega$cm and $T_{0}=$56 K.
The large deviation from the linearity is not discernible in $\rho$ (T) of samples with $x \geq$ 1.0 partially due to the contamination by ZrPt.

\begin{figure}
\begin{center}
\includegraphics[width=12cm]{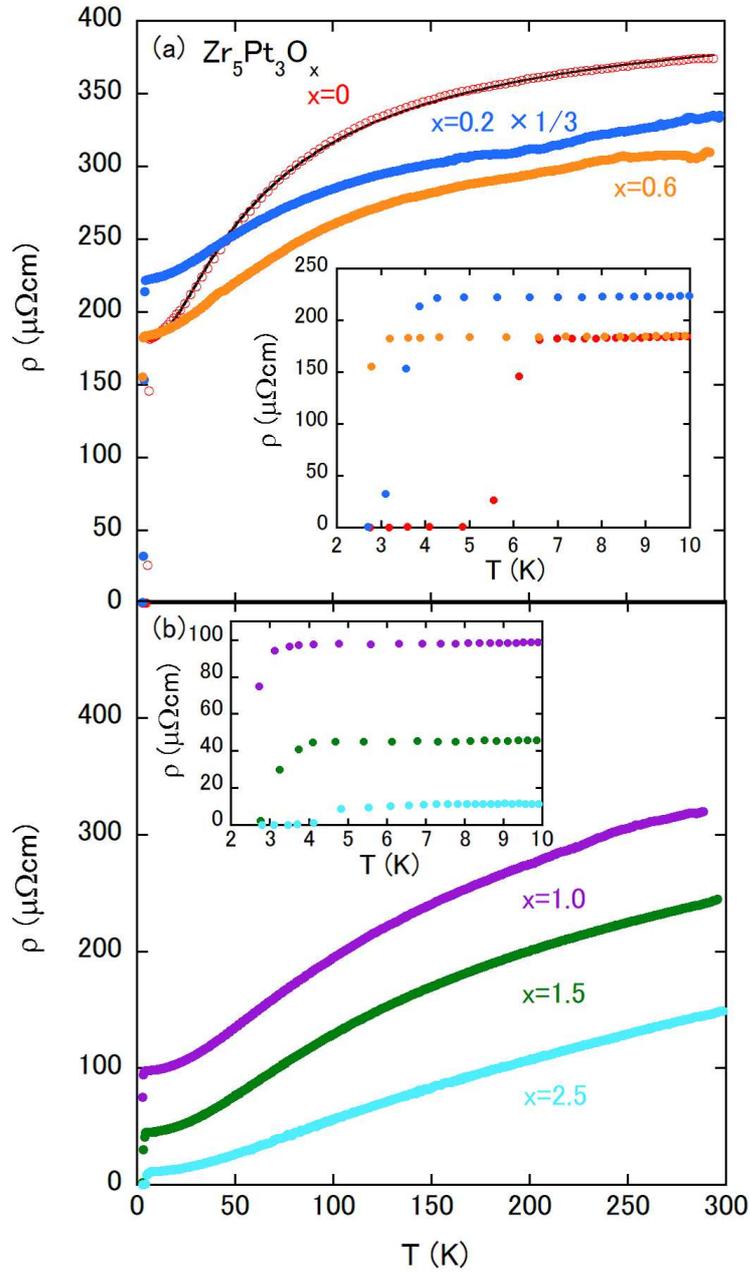}
\end{center}
\caption{(a) Temperature dependence of $\rho$ for Zr$_{5}$Pt$_{3}$O$_{x}$ with $x$=0, 0.2 and 0.6. $\rho$ (T) of Zr$_{5}$Pt$_{3}$O$_{0.2}$ is reduced to 1/3 of original scale. The solid curve is $\rho$ (T) calculated by eq.(1). The inset shows the low temperature part of each $\rho$ (T). (b) Temperature dependence of $\rho$ for Zr$_{5}$Pt$_{3}$O$_{x}$ with $x$=1.0, 1.5 and 2.5. The inset shows the low temperature part of each $\rho$ (T).}
\label{f6}
\end{figure}

Zr$_{5}$Pt$_{3}$O$_{x}$ with $x$= 0.2 and 0.6 can be regarded as the almost single-phased Ti$_{5}$Ga$_{4}$-type with the 2b site partially filled by oxygen atoms.
The oxygen addition up to $x$=0.6 systematically reduces $T_{c}$.
The similar systems of Nb$_{5}$Ir$_{3}$O and Zr$_{5}$Sb$_{3}$O also exhibit some interstitial-atom concentration dependences of $T_{c}$\cite{Zhang:npjQM2017,Lv:PRB2013}.
The reported dependences, combined with the variations of lattice parameters by atom additions, are summarized in Table 3.
In Nb$_{5}$Ir$_{3}$O, $T_{c}$ enhancement occurs with the addition of oxygen atoms, which shrinks $c$ and expands $a$.
The expansion of $a$ would bring octahedral Nb2$_{6}$ and Nb1 atomic chains closer to a one-dimensional system, and the shrinkage of $c$ leads to a shorter Nb-Nb distance.
These might cause the $T_{c}$ enhancement.
On the other hand, Zr$_{5}$Sb$_{3}$O and Zr$_{5}$Pt$_{3}$O$_{x}$ ($x\leq$0.6) show the reduction of $T_{c}$ by adding oxygen atoms.
In Zr$_{5}$Pt$_{3}$O$_{x}$ ($x\leq$0.6), both $a$ and $c$ decrease by oxygen addition, while the opposite trends are reported \cite{Lv:PRB2013,Robteutscher:ZMET1965} in Zr$_{5}$Sb$_{3}$O, suggesting the weak correlation between lattice parameters and  $T_{c}$.
In Zr compounds with the Mn$_{5}$Si$_{3}$-type structure, for example, the superconductivity appears\cite{Li:NJP2018} by substituting Ru into the Ge site in Zr$_{5}$Ge$_{3}$.
The substitution of atoms would be effective for the enhancement (appearance) of superconductivity in Zr based Mn$_{5}$Si$_{3}$ or Ti$_{5}$Ga$_{4}$-type structure.

\begin{table}[t]
\caption{Effect of interstitial atoms on superconducting properties and lattice parameters in Nb$_{5}$Ir$_{3}$O, Zr$_{5}$Sb$_{3}$O and Zr$_{5}$Pt$_{3}$O$_{x}$ ($x \leq$ 0.6). $a_{(p)}$ ($c_{(p)}$) and $T_{c(p)}$ are $a$ ($c$) and $T_{c}$ of parent compound (Nb$_{5}$Ir$_{3}$, Zr$_{5}$Sb$_{3}$ and Zr$_{5}$Pt$_{3}$), respectively. $+$ and $-$ mean the positive and the negative variation of $\Delta a$=$a$-$a_{(p)}$, $\Delta c$=$c$-$c_{(p)}$ and $\Delta T_{c}$=$T_{c}$-$T_{c(p)}$, respectively. The same information for Zr$_{5}$Ge$_{3-x}$Ru$_{x}$ is listed for the comparison.}
\label{t3}
\begin{tabular}{ccccc}
\hline
Compound & $\Delta a$=$a$-$a_{(p)}$ & $\Delta c$=$c$-$c_{(p)}$ & $\Delta T_{c}$=$T_{c}$-$T_{c(p)}$ & Reference \\
\hline
Nb$_{5}$Ir$_{3}$O & + & - & + & \cite{Zhang:npjQM2017}  \\
Zr$_{5}$Sb$_{3}$O & + & + & - & \cite{Lv:PRB2013,Robteutscher:ZMET1965} \\
Zr$_{5}$Pt$_{3}$O$_{x}$ ($x \leq$ 0.6) & - & - & - & this work  \\
Zr$_{5}$Ge$_{3-x}$Ru$_{x}$ & - & - & + & \cite{Li:NJP2018} \\
\hline
\end{tabular}
\end{table}

We discuss here the $T_{c}$ enhancement in Zr$_{5}$Pt$_{3}$O$_{x}$ with $x\geq$ 1.0, in which the superconducting Zr$_{5}$Pt$_{3}$O$_{\sim 0.5-0.6}$ phase coexists with ZrO$_{2}$ and ZrPt phases. 
Once the phase decomposition occurs at $x$=1.0, the oxygen content is reduced to approximately 0.45.
The increase of $x$ from 1.0 again gradually adds oxygen atoms.
Contrary to the above mentioned results of Zr$_{5}$Pt$_{3}$O$_{x}$ ($x \leq$ 0.6), the oxygen addition causes the $T_{c}$ enhancement.
As shown in Figs.\ 3(a) and 3(b), $a$, $c$ and $c/a$ might have some correlation with $T_{c}$ of samples with $x\geq$ 1.0; the expansion of both $a$ and $c$ with increased $c/a$-ratio possibly leads to the enhancement of $T_{c}$.
In order to elucidate the correlation between lattice parameters and $T_{c}$, we prepared a sample with the starting composition of Zr$_{4.9}$Pt$_{3.1}$O$_{0.67}$.
The compound, containing ZrPt impurity phase detected as the brighter image in Fig.\ 7, shows $a$, $c$ and $c/a$-ratio, which are between those of the sample with $x$=1.5 and $x$=2.5 (see Table 1).
Nonetheless $T_{c}$ of Zr$_{4.9}$Pt$_{3.1}$O$_{0.67}$ is lower than that of the sample with $x$=1.5 as demonstrated by Fig.\ 8.
This result suggests that the degree of $T_{c}$ enhancement is not determined only by the lattice parameters.
The important metallographic aspect is that, as $x$ is increased, the microstructure changes at $x \geq$1.0, showing the $T_{c}$ enhancement (see also Figs.\ 4(d) to 4(f)).
However, the microstructure of Zr$_{4.9}$Pt$_{3.1}$O$_{0.67}$ as shown in Fig.\ 7 is largely different from those of Zr$_{5}$Pt$_{3}$O$_{x}$ ($x \geq$1.0).
The synthesis with the oxygen content, exceeding the solubility limit, has changed a material's microstructure, which would play an important role for the $T_{c}$ enhancement.
We note here that in Sr$_{2}$RuO$_{4}$ or Ir a change of microstructure also enhances $T_{c}$\cite{Maeno:PRL1998,Matthias:Science1980}.
The eutectic Sr$_{2}$RuO$_{4}$ sample shows a lamellar pattern of Ru metal\cite{Maeno:PRL1998}.
Although the lamellar pattern does not affect the lattice parameters of Sr$_{2}$RuO$_{4}$, $T_{c}$ increases from 1.5 K to 3 K.
Ir with a small amount of YIr$_{2}$ forms a lamellar pattern and possesses a small lattice mismatch, leading to the strain-induced lattice softening\cite{Matthias:Science1980}.
This softening contributes to the $T_{c}$ enhancement from 0.1 K to 2.7 K. 

Another scenario of $T_{c}$ enhancement is a composition effect.
For example, if a ternary compound possesses a certain homogeneity range in its ternary phase diagram, $T_{c}$ or magnetic ordering temperature frequently depends on the composition of compound.
Such composition effect is reported for superconducting\cite{Ishikawa:PRB1983} CeCu$_{2}$Si$_{2}$ or antiferromagnetic compound\cite{Kitagawa:JPSJ1999} Nd$_{3}$Pd$_{20}$Ge$_{6}$. 
In CeCu$_{2}$Si$_{2}$, the lattice parameters do not exhibit clear composition dependence.
On the other hand, Nd$_{3}$Pd$_{20}$Ge$_{6}$ shows lattice parameters strongly depending on the composition.
Although the composition of Zr$_{5}$Pt$_{3}$O$_{0.6}$ with $T_{c}=$3.2 K is not so different from that of Zr$_{5}$Pt$_{3}$O$_{x}$ phase in Zr$_{5}$Pt$_{3}$O$_{2.5}$ with $T_{c}=$4.8 K (see Table 2), more precise study of composition effect might be needed.

\begin{figure}
\begin{center}
\includegraphics[width=9cm]{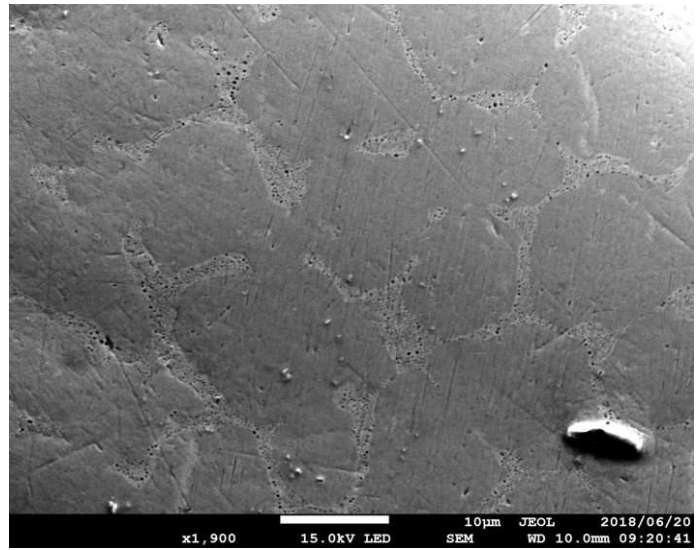}
\end{center}
\caption{Back-scattered electron (15 keV) image of Zr$_{4.9}$Pt$_{3.1}$O$_{0.67}$.}
\label{f7}
\end{figure}

\begin{figure}
\begin{center}
\includegraphics[width=10cm]{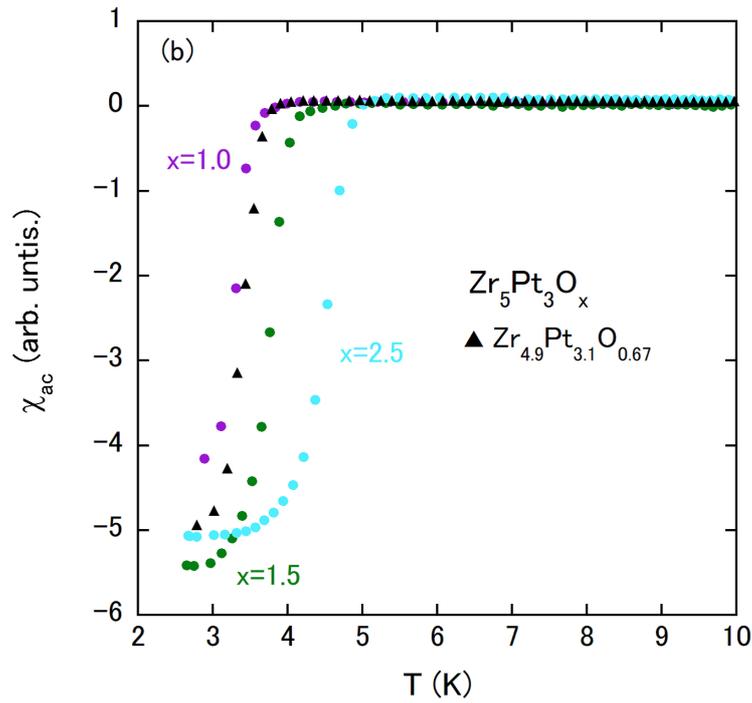}
\end{center}
\caption{Temperature dependence of $\chi_{ac}$ of Zr$_{5}$Pt$_{3}$O$_{x}$ with $x$=1.0, 1.5 and 2.5 and Zr$_{4.9}$Pt$_{3.1}$O$_{0.67}$.}
\label{f8}
\end{figure}

\section{Conclusions}
We have investigated the oxygen-content dependence of $T_{c}$ in Zr$_{5}$Pt$_{3}$O$_{x}$ by measuring $\chi_{ac}$ and $\rho$, combined with the metallographic study.
Single phase is confirmed for each sample with $x$=0 and 0.2.
A small amount of ZrO$_{2}$ appear in Zr$_{5}$Pt$_{3}$O$_{0.6}$.
Thus the oxygen content is limited to approximately 0.6.
With further increasing $x$ above 1.0, corresponding to the full occupancy of oxygen atoms, Zr$_{5}$Pt$_{3}$O$_{x}$ shows the decomposition into Zr$_{5}$Pt$_{3}$O$_{\sim 0.5-0.6}$, ZrPt and ZrO$_{2}$. 
The FE-SEM observation has confirmed the change of microstructure at $x \geq$1.0.
The parent Zr$_{5}$Pt$_{3}$ shows the superconductivity at $T_{c}$ of 6.4 K, which is decreased to 3.2 K as the oxygen content is increased to 0.6.
The crystallographic consideration is presented by surveying the results of Nb$_{5}$Ir$_{3}$O, Zr$_{5}$Sb$_{3}$O and Zr$_{5}$Ge$_{3-x}$Ru$_{x}$.
For Zr-based Mn$_{5}$Si$_{3}$-type compounds, oxygen addition would reduce $T_{c}$, irrespective of the oxygen-content dependence of lattice parameters.
The atom substitution like in Zr$_{5}$Ge$_{3-x}$Ru$_{x}$ might be effective for the enhancement of superconductivity.
In Zr$_{5}$Pt$_{3}$O$_{x}$ ($x\geq$ 1.0) showing the change of microstructure, slight enhancement of $T_{c}$ with increasing $x$ is observed.
The fact would be correlated with the change of microstructure and/or the composition effect.

\ack
J.K. is grateful for the financial support provided by the Comprehensive Research Organization of Fukuoka Institute of Technology.

\end{document}